*Intergenerational AI Literacy in Korean Immigrant Families: Interpretive Gatekeeping Meets Convenient Critical Deferment*

JEONGONE SEO, Rutgers University, USA[1]

RYAN WOMACK, Rutgers University, USA

TAWFIQ AMMARI, Rutgers University, USA

**Abstract** As artificial intelligence (AI) becomes deeply integrated into family life, immigrant families must navigate unique intergenerational, linguistic, and cultural challenges. This study examines how Korean immigrant families in the United States negotiate the use of AI tools such as ChatGPT and smart assistants in their homes. Through 20 semi-structured interviews with parents and teens, we identify two key practices that shape their engagement: interpretive gatekeeping, where parents mediate their children's AI use through a lens of cultural and ethical values, and convenient critical deferment, where teens strategically postpone critical evaluation of AI for immediate academic and social utility. These intertwined practices challenge conventional, skills-based models of AI literacy, revealing it instead as a dynamic and relational practice co-constructed through ongoing family negotiation. We contribute to information science and HCI by offering a new conceptual extension for intergenerational AI literacy and providing design implications for more equitable, culturally attuned, and family-centered AI systems.

# 1. Introduction

Artificial intelligence (AI) tools—ranging from generative platforms like ChatGPT to smart assistants such as Alexa—are increasingly embedded in the everyday lives of youth and families. While research has examined the integration of AI in educational and domestic settings, relatively little attention has been paid to how families from immigrant backgrounds negotiate AI use across generational, linguistic, and cultural boundaries. For immigrant families, AI tools can simultaneously offer empowerment and exclusion, depending on how they align with familial values, digital competencies, and language fluencies.

AI literacy is conventionally defined as a set of individual competencies, encompassing the ability to understand how AI systems work, critically evaluate AI-generated outputs, and use these tools ethically and effectively (e.g., Long & Magerko, 2020; Ng et al., 2021). This perspective often frames literacy as a universal collection of skills that an individual must acquire to be considered "literate." While valuable, such an individualistic and skills-based framework is often insufficient for understanding how AI literacy is enacted within complex sociocultural contexts like immigrant families. It tends to overlook the relational dynamics, linguistic divides, and intergenerational value negotiations that profoundly shape how technology is understood, adopted, or resisted in everyday life.

This study uses a relational approach grounded in Information Practice Theory, which frames information engagement not merely as skill acquisition but as socially and culturally situated meaning-making (Savolainen, 1995; Lloyd, 2010). We also build on perspectives from Distributed Cognition, recognizing that knowledge and literacy are not located in a single person's mind but are distributed across family members, technologies, and cultural norms (Pea, 1993; Druga et al., 2022). Finally, Information Worlds Theory (Jaeger & Burnett, 2010) provides a powerful lens to analyze how families negotiate the conflicting information norms and values that arise from their unique position bridging different generational, cultural, and institutional worlds.

---

[1] Corresponding author: joh.seo@rutgers.edu

These theoretical lenses help illuminate the mechanisms through which immigrant families engage with AI not just as users, but as meaning-makers navigating linguistic hierarchies and cultural tensions. In particular, parental strategies for regulating and contextualizing their children's AI use draw attention to a form of mediation that goes beyond access control or usage rules. Prior work on parental mediation has documented restrictive, monitoring, and active forms (Livingstone & Helsper, 2008) but less is known about how immigrant parents infuse these practices with cultural, religious, and moral values in response to AI's perceived epistemic risks. Given this gap, we ask:

**RQ1. How do Korean immigrant parents culturally and ethically shape their children's AI use?**

While parents navigate AI from a stance of cautious evaluation, teens often approach it from a position of pragmatic adaptation. Building on emerging research around adolescent engagement with generative AI (Choi et al., 2025; Han et al., 2024), we observe that youth do not abandon critical thinking entirely but instead defer it to optimize efficiency. This behavior—what we term convenient critical deferment—reflects both contextual constraints and a strategic mindset. However, few studies have examined how immigrant teens balance AI's utility against concerns about voice, authorship, and accuracy, particularly under academic and linguistic pressure. To address this, we ask:

**RQ2. How do Korean immigrant teens adopt AI at the intersection of academic, linguistic, and time pressures?**

AI technologies do not function in isolation within the household. Instead, our findings reveal that interactions with AI often become moments of family negotiation, cooperation, or subtle tension (Druga et al., 2022). These interactions give rise to a unique, family-specific ecosystem of AI literacy, shaped by teen-driven digital mediation and parent-led ethical framing. This dynamic reflects the concept of Joint Media Engagement (Takeuchi & Stevens, 2011), which emphasizes shared technology use as an interactive and co-constructed process rather than a series of parallel actions. Within this framework, AI becomes a shared point of exploration: teens support the technical engagement, while parents provide cultural and ethical context. Together, these roles form a distributed and evolving literacy practice. However, little is known about how these practices intersect—whether they foster mutual learning or reinforce existing power asymmetries. This prompts us to ask:

**RQ3. How do parents and teens co-construct AI literacy within immigrant family settings?**

This study moves beyond a static view of literacy as individual competence. We propose to conceptualize intergenerational AI literacy not as a stable set of skills held by individuals, but as a distributed, negotiated, and relational practice. This approach allows us to see literacy as something that is dynamically co-constructed within the family unit, emerging from the tensions and collaborations between members. To build this framework, we draw upon several key theoretical pillars from information science.

# 2. Related Literature

To situate our analysis of AI literacy within Korean immigrant families, we draw on three intersecting strands of scholarship. First, we engage with literature on parental mediation and interpretive gatekeeping to examine how immigrant parents shape their children's technology use through culturally grounded, ethically motivated practices. Second, we incorporate emerging research on adolescent pragmatism in AI engagement to frame teens' strategic use of AI as a situated response to academic and linguistic pressures. Third, we turn to studies of intergenerational co-construction and distributed cognition to understand how family members collaboratively negotiate AI literacy through dialogic interactions and shifting roles. Together, these literatures illuminate how AI literacy emerges not as an individual skillset but as a relational and culturally embedded practice, co-constructed through ongoing family negotiation. Yet this motivational account remains necessarily incomplete. Zhang et al.'s (2025) focus on a monolingual population overlooks the cultural and linguistic particularities that shape literacy practices in immigrant households. In contrast, this manuscript complicates conventional notions of literacy by highlighting how Korean-American families navigate migration-related stressors, including bilingual power reversals and protective gatekeeping—dynamics that remain invisible in studies that do not account for such sociolinguistic and cultural complexities.

## 2.1 Parental Mediation and Interpretive Gatekeeping

The concept of gatekeeping, which originates from communication studies to describe how information is filtered for a group (e.g., Shoemaker, 1991), offers a foundational lens for understanding parental mediation. While parental mediation has traditionally been typologized as restrictive, monitoring, or active (Livingstone & Helsper, 2008). However, recent research indicates that in immigrant contexts, mediation strategies function as interpretive acts grounded in values, religion, and developmental intentions (Bhatti et al., 2025; Beneteau et al., 2020; Jang et al., 2025). These studies suggest that parents do not merely regulate access to technology but repurpose AI as a tool for moral guidance, language refinement, and early education.

Religious and cultural values often shape content selection, where screen media devices are used purposefully to reinforce spiritual knowledge and ethical behavior (Bhatti et al., 2025). Smart assistants also support parental scaffolding by prompting children to adjust their speech for better comprehension, thus facilitating language learning in daily routines (Beneteau et al., 2020). This aligns with Jang et al. (2025), who show that semantically rich, socially embedded conversations during routines like mealtime significantly influence syntactic development. Such learning environments illustrate how interpretive gatekeeping operates as a distributed literacy practice, embedded in the flow of family life.

Immigrant parents also draw on cultural brokers—such as school liaisons—to help navigate digital and institutional information systems. These liaisons not only translate content but mediate it in culturally resonant ways that enhance comprehension and trust (Wong-Villacres et al., 2019). Their role reflects the concept of information grounds, where information exchange is socially facilitated and contextually adapted (Fisher & Naumer, 2006), aligning with Savolainen's (2008) view of information practices as socially and culturally established ways of engaging with information in everyday contexts. As Lloyd (2010) argues, such practices represent a form of situated information literacy, understood as embodied information work situated in everyday life. This perspective is further exemplified in studies of marginalized groups, such as refugee youth, whose information literacy practices are deeply embedded in their daily navigations of new social and informational landscapes (Lloyd & Wilkinson, 2016).

Taken together, these studies highlight that interpretive gatekeeping is not merely about shielding children from digital risks but actively shaping meaning through cultural, moral, and developmental lenses. Recent work by Seo and Ammari (2025) similarly conceptualizes selective disengagement among older Korean immigrants not as digital deficiency, but as "pragmatic disengagement"—a strategic, culturally grounded refusal of emotionally taxing or linguistically misaligned content. Their framework helps situate interpretive gatekeeping as not just a parental behavior, but a broader mode of protective information practice shaped by cultural norms and emotional boundaries.

## 2.2 Teen Pragmatism and Convenient Critical Deferment

Teen engagement with AI systems reflects a complex balance between critical awareness and practical expediency. Rather than uniformly interrogating AI outputs, immigrant teens often deploy a context-sensitive strategy we term convenient critical deferment. This involves suspending judgment in favor of efficiency, particularly when under time pressure, facing language barriers, or working on low-stakes tasks (Choi et al., 2025; Han et al., 2024; Saxena et al., 2025).

Studies have documented how students copy-paste AI-generated content without editing, rely on heuristics like cross-checking with Google or classmates, and toggle between skepticism and acceptance depending on perceived task importance (Choi et al., 2025; Han et al., 2024). These strategies indicate not a lack of criticality but an effort to balance agency and cognitive burden.

Linguistic pragmatism also plays a key role. Bilingual youth often code-switch to ease communication with AI systems and flexibly assign trust depending on whether information is factual or emotionally nuanced (Chen, 2024; Cihan et al., 2022; Lee, J.Y., et al., 2025). Migrant teens further use AI tools to simulate culturally specific scenarios and explore personal identity in anticipation of relocation (Lee, S., et al., 2025).

This pattern aligns with findings that users are often tolerant of AI imperfections in creative or exploratory tasks and accept ambiguity as part of the digital experience (Saxena et al., 2025). Duolingo, for instance, embraces error-prone interaction by combining AI suggestions with human-curated oversight, a model that teens intuitively adapt to in informal settings.

Parental influence also shapes teen pragmatism. Sun et al. (2021) show that as children mature, parents shift from restrictive control to educational trust-building—encouraging critical reflection while acknowledging autonomy. Our data reflect this transition, as teens cite both inherited caution and conscious flexibility in their AI habits.

Convenient critical deferment thus captures how youth pragmatically manage digital complexity, not by rejecting criticality, but by staging it in ways that fit their constraints and goals (Savolainen, 2008).

## 2.3 Intergenerational Co-Construction of AI Literacy

AI literacy in immigrant families is not transferred unilaterally from adult to child; rather, it is co-constructed through reciprocal learning, shifting roles, and dialogic interactions. Families jointly explore, test, and reflect on AI in ways that challenge traditional generational hierarchies and create opportunities for shared meaning-making (Druga et al., 2022; Livingstone & Helsper, 2008; Davis et al., 2019).

Studies show that teens frequently act as digital mentors, while parents contribute contextual or ethical frames—resulting in dynamic learning partnerships (Davis et al., 2019; Druga et al., 2022). Within these exchanges, smart devices become embedded in shared routines such as games, trivia, or language play, facilitating co-use and collaborative interpretation (Ammari et al., 2019; Oewel et al., 2023; Beneteau et al., 2020).

Emotional regulation and relational insight also emerge through AI-mediated communication. Shen et al. (2025) found that AI-generated conversation summaries and prompts enable parents to track emotional nuance without full co-viewing, supporting stronger connection. Claggett et al. (2025) observed that exposure to prosocial message suggestions enhances conversational tone and cooperative dynamics, even when prompts are not actively used.

Cultural transmission is further supported by narrative interaction. Liaqat et al. (2022) and Shen et al. (2024, 2025) demonstrate how storytelling tools bridge generational and linguistic gaps, prompting intergenerational empathy and reflection. These platforms offer third spaces where families can renegotiate shared identities and construct hybrid meanings.

Seo and Ammari (2025) emphasize the role of "interdependent navigation"—a condition in which older adults' participation in digital life is scaffolded by their children, caregivers, and social environments. This concept expands the notion of co-construction by highlighting how relational asymmetry, emotional labor, and cultural responsibility shape what literacy means across generations.

AI literacy emerges as a distributed process of knowledge construction, shaped by language, trust, humor, and storytelling. The home, in this view, becomes an information ground (Fisher & Naumer, 2006) and a relational infrastructure for culturally situated AI learning, reflecting Savolainen's (1995) emphasis on everyday life contexts as central to information practices. This perspective is further enriched by Lloyd's (2010) work on information literacy as an embodied, social practice enacted within these everyday landscapes.

# 3. Methodology

To investigate how Korean immigrant families engage with AI technologies in everyday life, we conducted a qualitative study using semi-structured interviews with parents and teens in the New York metropolitan area. This section details our research design, participant recruitment strategy, data collection procedures, and thematic analysis approach.

## 3.1 Research Design

This qualitative study examined how Korean immigrant families in the New York metropolitan area engage with artificial intelligence (AI) tools in their daily and educational lives. A semi-structured interview method was employed to capture intergenerational perspectives on AI use, digital literacy, language barriers, and trust in technology. The research design was informed by a culturally grounded and family-based approach, with particular attention to bilingual and cross-generational dynamics.

## Participant Recruitment

Participants were recruited via purposive and snowball sampling strategies through outreach to Korean community organizations, churches, Korean language schools, and online community forums. Recruitment flyers were provided bilingually (Korean and English), and each participating family was offered a $25 incentive per individual.

Inclusion criteria included:

- One or more Korean immigrant parents or guardians residing in the New York metropolitan area;
- One adolescent aged 14–18 within the same household;
- Prior use or awareness of AI-powered tools (e.g., ChatGPT, Google Translate, AI-based learning apps).

## 3.2 Data Collection

Interviews were conducted from March to April 2025, either in person, via Zoom, or by phone, depending on participant preference and availability. Interviews lasted 40–60 minutes and were conducted in the participant's preferred language (Korean or English). All sessions were audio-recorded with informed consent, except where participants opted for detailed note-taking.

Interview guides were tailored by role:

- Parents were asked about their children's AI use, perceived benefits and risks, language-related challenges, and family-based technology practices;
- Teens discussed their own use of AI tools, trust in AI outputs, language preferences, and family communication around technology.

Although the study was designed to include matched parent-teen pairs, only a subset of participants were dyadically matched due to logistical challenges. Given that not all participants were dyadically matched, our analysis focuses on identifying overarching patterns within each generational group (parents and teens) first, before exploring the interactive dynamics primarily through the data from matched pairs.

## 3.3 Data Analysis

A qualitative thematic analysis was employed. Interviews were transcribed verbatim in their original language and reviewed for accuracy. Transcripts were then coded using a mixed inductive-deductive approach. Key thematic categories included:

- Language-based differences in AI usability;
- Intergenerational perceptions of AI trustworthiness;
- Role reversal in digital instruction and mediation;
- Cultural concerns regarding AI's educational and ethical impacts.

The coding process was conducted iteratively, with intercoder reliability checks and memo-writing to refine interpretations. Cross-generational and bilingual contrasts were emphasized to uncover patterns specific to immigrant family contexts.

Table 1. Participant Demographics and Family Links. Note: In some cases, either parent or teen declined or was unable to participate despite expressed interest. This is noted in the table.

| Participant ID | Role | Gender | Age Group | Interview Language | Education Level | Family Link (Child or Parent ID) |
|---|---|---|---|---|---|---|
| P1 | Parent | Female | Mid 50s | Korean | Bachelor's | T5 |
| P2 | Parent | Male | Late 40s | Korean | Bachelor's | T2 |
| P3 | Parent | Female | Mid 50s | Korean | Bachelor's | T6 |
| P4 | Parent | Female | Late 40s | Korean | Bachelor's | T2 |
| P5 | Parent | Male | Mid 50s | Korean | Bachelor's | T6 |
| P6 | Parent | Male | Mid 50s | Korean | Ph.D. | Did not interview |
| P7 | Parent | Male | Late 50s | Korean | Ph.D. | T7 |
| P8 | Parent | Female | Late 40s | Korean | Ph.D. | T3 |
| P9 | Parent | Male | Early 60s | English | Bachelor's | T5 |
| P10 | Parent | Female | Late 50s | Korean | Ph.D. | Did not interview |
| T1 | Teen | Female | Late teens | English | High School | Parent not interviewed |
| T2 | Teen | Female | Late teens | English | High School | P2, P4 |
| T3 | Teen | Female | Late teens | English | High School | P8 |
| T4 | Teen | Male | Late teens | Korean | High School | Parent not interviewed |
| T5 | Teen | Male | Late teens | English | High School | P1, P9 |
| T6 | Teen | Male | Late teens | English | High School | P3, P5 |
| T7 | Teen | Female | Late teens | English | High School | P7 |
| T8 | Teen | Female | Late teens | English | High School | Parent not interviewed |
| T9 | Teen | Female | Late teens | English | High School | Parent not interviewed |
| T10 | Teen | Male | Late teens | Korean | High School | Parent not interviewed |

Ethical Considerations

The study received Institutional Review Board (IRB) approval from Rutgers University (Protocol #Pro2024002241). All participants completed informed consent or assent processes in their preferred language. Confidentiality was ensured through pseudonym assignment, encrypted storage of transcripts, and deletion of audio files after transcription. De-identified data will be stored securely for six years in accordance with IRB policies.

# 4. Findings

In the following section, we present our findings across three interrelated thematic areas that illuminate the dynamics of intergenerational AI literacy in Korean immigrant families. First, we examine interpretive gatekeeping, highlighting how parents enact culturally and ethically grounded forms of mediation to regulate their children's AI use. This theme reveals how parental practices are shaped by concerns over epistemic responsibility, cultural preservation, and linguistic accessibility. Second, we explore the practice of convenient critical deferment among teens, demonstrating how youth strategically suspend critique to maximize efficiency under academic and linguistic pressures. This section focuses on teens' pragmatic engagement with AI tools and the contextual constraints that shape their choices. Third, we analyze intergenerational mediation and co-construction, which captures how families collectively navigate AI through bilingual tensions, role reversals, and moments of dialogic learning. Together, these three themes offer a holistic view of AI literacy as a negotiated, relational, and culturally situated practice within immigrant households.

## 4.1. Parental Interpretive Gatekeeping: Ethical Oversight, Cultural Anchoring, and Linguistic Mediation

This section examines how Korean immigrant parents practice interpretive gatekeeping—a culturally and ethically informed form of digital mediation. Unlike restrictive models focused on control, these parents act as evaluators, assessing AI tools for their fit with family values, educational standards, and language needs. We find that their aim is to guide and contextualize AI use, not to prohibit it.

### 4.1.1. Defining Interpretive Gatekeeping

Across the interviews, parents acknowledged their children's increasing independence in engaging with AI tools but still felt compelled to shape how such engagement unfolds. Their concern was not about banning access outright, but about instilling evaluative thinking and moral caution.

As P2 noted, "I told her about services like ChatGPT or Grok, and when I see her using them when necessary, I tell her to think while using such tools and to check if the answer is actually right." Similarly, P3 added, "I don't prohibit [that she uses AI tools], but I tell her to only use AI as a reference. Not to fully trust it." These perspectives reflect a shared desire to embed intentionality and epistemic responsibility into AI use, especially when teens are using it in solitary or school-driven contexts. Other parents echoed this cautionary stance. P4 explained, "We don't particularly provide AI tools at home, but my child uses AI-based materials provided at school. I just tell her not to believe everything at face value." This suggests an openness to institutionalized use paired with a reluctance to fully endorse unsupervised access.

P8 offered a more direct interventionist role: "My daughter showed me an essay she generated using AI, and I made her rewrite parts of it. I told her she can use it as a base, but she needs to add her own thoughts." These comments show that parents were not rejecting AI wholesale but acting as value-oriented moderators—guiding their children's digital behaviors toward intentional and reflective use.

### 4.1.2. Cultural and Educational Values in Tension

A key concern for parents was the tension between AI's convenience and traditional Korean values emphasizing perseverance, deep learning, and sustained effort. They feared AI might encourage children to shortcut the learning process.

As P9 put it, "AI saves time, but I worry it leads kids to skip the effort of learning," expressing concern not about misinformation, but about eroding academic discipline. P10 similarly noted, "Korean expressions often carry subtle meanings that AI can't fully capture," highlighting worries about a loss of cultural and cognitive depth. For P6, the pace of AI clashed with traditional, intentional education: "Children these days want everything fast... I don't want my child to lose that [depth] just because a machine gives a quick answer."

Through interpretive gatekeeping, parents seek to preserve cultural learning norms while cautiously integrating digital tools into their children's lives.

### 4.1.3. Language as a Site of Cultural Negotiation

Language was a key site of interpretive gatekeeping. While most parents used AI in Korean, many found its outputs awkward or unnatural, affecting both their trust in the tool and their support for its use by their children.

P1 noted, "It would be better if the Korean was more natural… the translation is really awkward," reflecting both linguistic dissatisfaction and a sense of cultural misrecognition. P5 said, "I try to use it in English, but it's hard. So I ask my son sometimes," revealing a reversal of traditional roles, where children act as linguistic and epistemic mediators. However, P7, who is fluent in English, observed, "It feels like we're locked out of something because the system wasn't made for us," underscoring how language barriers in digital tools amplify feelings of exclusion.

### 4.1.4. Trust, Skepticism, and Conditional Engagement

A core dimension of interpretive gatekeeping was the parents' conditional trust in AI. While many acknowledged its usefulness, they approached it with skepticism—particularly concerning how information is generated and whether it aligns with their epistemic standards.

P3 remarked, "AI gives the answer, but how do we know if it's really correct?" This skepticism extended beyond factual correctness to broader concerns about transparency and cultural fit. P4 added, "It's not about accuracy alone. It's about whether it makes sense in our context." Several parents emphasized the need for verification. P8 noted, "Sometimes the answer is totally off. I tell my daughter to always double-check with other sources or ask a teacher." P9 added, "English responses are more accurate, but Korean still needs a lot of improvement. That's why I don't let my child rely solely on AI." Tech-savvy parents like P7, who holds a Ph.D., commented, "AI is helpful for summaries or basic tasks, but it shouldn't replace thinking. I remind my son: the tool is only as smart as the person using it."

This dialogic stance—encouraging reflection, verification, and context-based understanding—was central to how parents exercised interpretive gatekeeping. Their goal was not to control but to cultivate discernment.

## 4.2. Teens' Convenient Critical Deferment: Pragmatism, Temporal Deferral, and Intergenerational Navigation

This section explores how Korean immigrant teens engage AI with strategic pragmatism. Rather than rejecting critical thought, they temporarily defer judgment to gain AI's convenience—a practice we term *convenient critical deferment*. Here, critique is delayed, not discarded, enabling AI to serve as a practical cognitive aid. Their approach reflects a situated literacy shaped by academic pressure, language, and family dynamics.

### 4.2.1. Strategic and Frequent Use of AI Tools

Teen participants reported high-frequency, goal-oriented use of AI tools, particularly in academic contexts. They used AI for writing support, translation, brainstorming, and comprehension. However, this use was rarely framed as dependence. Instead, teens described AI as a functional assistant or productivity enhancer.

Teens described AI tools as seamlessly integrated into their study routines. T6 said, "AI tools are really useful… they help me generate ideas, summarize, and check grammar quickly," while T9 called ChatGPT "a smart helper" and emphasized learning how to prompt effectively. T3 added, "I use it all the time… especially when studying or trying to understand questions." This normalization suggests teens view AI as a routine aid, not a disruptive force. Still, their use was selective. As T5 noted, "AI should only be used for tedious parts… it helps make a foundation, and then I build on it." Teens thus used AI to support—not replace—their academic thinking.

### 4.2.2. Awareness of Limitations and Postponed Critique

Teens were aware of AI's limitations and practiced deferred critical thinking—using tools for convenience but verifying results afterward.

T2 said, "I feel more confident asking in English… but I go over the result before I submit anything," reflecting conditional trust. T8 was more cautious: "By using AI, I would be subjecting myself to limited thinking… I hesitate to use it unless I really need it," revealing skepticism about overreliance. T7 struck a balance saying that AI "speeds things up, but I know it's not perfect. I usually check with someone or rewrite parts."

These accounts challenge the idea that teens uncritically accept AI, showing instead a pattern of pragmatic use paired with reflective oversight.

### 4.2.3. AI as a Bridge in Resource-Constrained Environments

AI was often described as compensating for limited academic support. Especially for teens whose parents were less fluent in English or less familiar with school systems, AI served as a silent tutor or translator.

T6 shared, "It can be confusing sometimes because of the way teachers write. AI helps me understand the structure or meaning of the question." Similarly, T10 explained, "AI is very efficient for school. It helps me do my work faster, and I use it even for organizing my life." These quotes illustrate how AI filled gaps that traditional support systems could not address.

T9 expressed appreciation for AI's accessibility: "Sometimes I don't want to ask my parents or teachers because they're busy or don't get it. I just ask ChatGPT and try to figure it out." This indicates that AI was not only a technical aid but a relational buffer, reducing friction in intergenerational communication.

Still, some teens resisted reliance on AI. T8 said, "I hesitate to use AI unless I'm stuck. I try to do it myself first, but I know it's there when I need it." This measured approach reveals how AI was positioned within a broader ecology of self-regulation and learning goals.

### 4.2.4. Language Choice and Cultural Distance

Most teens reported using AI exclusively in English, even when they spoke Korean at home. English was perceived as the more functional, precise, and authoritative language in AI interactions.

T1 remarked, "I have never used AI tools in Korean, but I have used translation tools for Korean assignments. I usually use English because the content is better." T7 echoed this, noting, "I find it easier to put in English and for it to understand what I mean. Korean sometimes gives weird answers." T10 reflected on this linguistic alignment: "Personally, I use English more because my work and life are in the U.S. I use Korean sometimes to find specific cultural things, but not often." For many participants, English was the 'default' digital language, even if it was not their first language at home.

However, this preference created distance from their cultural and familial language practices. T9 noted, "When I need translations, the Korean sometimes doesn't make sense. I think it's trained more in English." This gap shaped how teens explained AI use to parents—often acting as informal translators or brokers.

Though not always explicitly recognized, this linguistic asymmetry subtly reinforced the idea that Korean was less compatible with digital authority. Teens rarely expressed frustration, but their linguistic choices signaled broader dynamics of cultural displacement.

### 4.2.5. Balancing Efficiency and Responsibility

Teens consistently managed a balance between leveraging AI's efficiency and retaining intellectual responsibility. They used AI not to replace effort but to redistribute it—saving time for ideation, revision, or other tasks.

T3 explained, "Honestly I know people are freaking out about AI taking over, but it's just a tool. You still have to do the thinking." This reflects a pragmatic but bounded use of AI, in which the user maintains epistemic agency. T6 summarized this dynamic well: "AI can help me get started, but I don't just copy-paste. I look over everything and change it to match my style." This awareness of voice and ownership suggests a form of digital maturity. Even when time pressures or language challenges led to heavier AI use, teens emphasized that final responsibility rested with them. As T7 stated, "It's helpful, but I know it's not a replacement. I always make sure it sounds like me."

In sum, teen participants practiced a nuanced form of digital literacy. Their "convenient critical deferment" was not ignorance, but a calculated response to academic demands, linguistic constraints, and the expectation of self-directed learning. They recognized AI's strengths and limits—and chose to engage it on their own terms.

## 4.3. Intergenerational Mediation and Co-Construction: Bilingual Tensions, Role Reversals, and Trust Negotiation

In bilingual Korean immigrant households, AI use reshaped generational roles. Parents approached it cautiously, shaped by cultural values and language barriers, while teens navigated it confidently through English interfaces. This imbalance sparked both friction and collaboration, leading to a co-construction of knowledge through informal yet meaningful exchanges about technology, trust, and expertise. This section examines how these dynamics played out across generations.

### 4.3.1. Teens as Digital Intermediaries

Teens often took on the role of digital intermediaries in their households, a position that emerged organically from their greater fluency with technology and English-language platforms. This role was typically informal, shaped by linguistic and digital asymmetries rather than explicit expectations. Many teens reported helping parents navigate apps, troubleshoot settings, or understand AI tools, effectively acting as informal tutors.

While some exceptions existed—such as tech-savvy parents who occasionally guided their children—these were less common. As T7 put it, "My mom likes to teach me traditional things, but I'm the one who explains how ChatGPT works." This reflection encapsulates the broader pattern: a shift in household expertise that fosters intergenerational learning while subtly challenging traditional norms of parental authority.

### 4.3.2. Intergenerational Trust, Language, and Co-Learning in

Trust in AI systems varied notably across generations, shaped by differences in language fluency, digital familiarity, and perceived credibility. Teens generally expressed functional confidence in AI, especially when using English. T1 explained, "AI tends to give relevant information. But I still check it," while T6 emphasized trust in English-language outputs. In contrast, parents remained more cautious—particularly toward Korean-language responses—voicing concerns about translation quality, cultural mismatch, and lack of clarity. As P1 remarked, "It would be better if the Korean was more natural. The translation is really awkward." These trust disparities were not simply personal—they reflected broader structural asymmetries in AI design, where English dominates and other linguistic preferences are deprioritized.

Language thus played a pivotal role in shaping digital access and authority. Teens preferred using AI in English, not just for ease but due to perceptions of higher accuracy. T2 noted that Korean outputs "give off answers that don't make sense," while T9 described Korean as a "backup language." Meanwhile, parents—particularly recent immigrants like T4—continued to rely on Korean, even as it constrained their interaction with AI. These patterns deepened epistemic divides: teens moved fluidly across platforms, while parents encountered friction, often deferring to their children's expertise.

As teens assumed interpretive authority, they framed their AI use as efficient and under control. T3 commented, "It's just a tool. You still must do the thinking," distancing themselves from fears of overreliance. Parents, however, viewed AI with more ambivalence, sometimes feeling excluded from digital routines. Still, rather than escalating into generational conflict, these differences often prompted quiet adjustments and learning opportunities.

Despite these asymmetries, many families experienced moments of co-learning. In these interactions, teens and parents collaborated to troubleshoot, explore, or interpret AI outputs—revealing relational dynamics embedded in technology use. T10 shared, "My mom sometimes asks me to check something in English… she tells me to slow down and explain why it says that." Similarly, T5 described how their parents' curiosity was piqued by observing ChatGPT use, leading to casual yet meaningful exchanges. T7 recalled testing AI's accuracy with their father by cross-checking answers with Wikipedia: "It was kind of fun."

These moments of co-engagement were often experiential rather than didactic. When teens led, their tone was typically cooperative rather than authoritative, fostering mutual reflection. Such role reversals challenged traditional norms of expertise and showed that AI literacy, rather than being an individual skill, could become a shared, intergenerational project.

### 4.3.3. Toward Family-Based AI Literacy

Despite moments of disconnect, the data revealed that AI literacy in immigrant households was not individual but distributed. Teens brought technical agility and linguistic confidence; parents contributed ethical reasoning and contextual caution. Together, they shaped a family-specific literacy practice.

While no participant explicitly described setting joint rules, their stories implied layered logic: teens deferred critique, parents insisted on reflection. When these orientations intersected—particularly through conversation or shared experimentation—families developed hybrid approaches to AI use.

T2 summarized this co-learning ecology well: "I use AI all the time, but my mom asks questions that make me think. Sometimes I just want to finish, but her questions actually help." In sum, Korean immigrant families navigated AI not as isolated individuals but as interconnected actors. Their literacy practices were shaped by intergenerational negotiation, linguistic complexity, and ethical concerns. Far from being a source of conflict alone, AI became a medium through which families redefined roles, shared knowledge, and co-constructed meaningful engagement with technology.

# 5. Discussion

This study's findings, centered on the concepts of *interpretive gatekeeping* and *convenient critical deferment*, reveal that AI literacy within Korean immigrant families is not a uniform or individual competency. Rather, it is a distributed, relational practice dynamically co-constructed through intergenerational, linguistic, and culturally grounded negotiations. In response to our research questions, the data illuminate how parents assert culturally and ethically driven oversight (RQ1), how teens strategically defer critique to manage academic and temporal pressures (RQ2), and how these practices interact to shape shared understandings of AI within the household (RQ3).

## 5.1. Challenging Existing Frameworks: From Individual Skills to Relational Practices

Our findings challenge the conventional understanding of AI literacy as merely an individual skill set. Grounded in Information Practice Theory (Savolainen, 1995), we acknowledge that everyday information seeking is deeply influenced by one's life context and habitual orientations. Expanding on this, Lloyd (2008) conceptualizes information literacy as a socially enacted and embodied practice, situated within particular domains of activity. In line with this trajectory, our study demonstrates that information practices within immigrant families are not solely shaped by access or usage, but also by culturally rooted acts of protection, emotional regulation, and linguistic negotiation. Lloyd and Wilkinson's (2016) research on refugee youth highlights how domestic spaces—such as the home—serve as critical sites for informal literacy development. Similarly, in our case, the Korean immigrant household emerges as a dynamic setting of intergenerational mediation, where knowledge is co-constructed amid structural and emotional constraints. Within this space, parents' cautious moderation—centered on education, language, and ethics—acts as a protective strategy to preserve cultural identity, resonating with Seo and Ammari's (2025) concept of "pragmatic disengagement."

Likewise, teens' convenient critical deferment reframes digital pragmatism not as lack of criticality, but as a situated response to the burdens of bilingualism, high academic stakes, and intergenerational expectations. This supports Choi et al. (2025) and Han et al. (2024), who argue that youth pragmatism is often an adaptive, context-sensitive strategy. Together, these practices illustrate a shift from a skills-based model of AI literacy to a fluid, contextually embedded practice negotiated across family roles and values. Our findings build on and, in key respects, complicate the analysis presented by Zhang et al. (2025). While their study offers valuable insight into the motivational dynamics behind family engagement with ChatGPT, our work expands this lens by examining how such motives are mediated through the cultural and linguistic specificities of immigrant households. In particular, we highlight the role of bilingual friction, commitments to cultural preservation, and intergenerational power negotiations in shaping these interactions. Through this reframing, we move beyond conceptualizing AI literacy as an aggregation of individual "needs," instead proposing it as a socially and culturally situated negotiation of authority across generations and languages.

## 5.2. Information Worlds in Conflict: Cultural Frictions and Role Reversals

Building on Jaeger and Burnett's (2010) Information Worlds Theory, we observed how Korean immigrant households become contested terrains where multiple information worlds coexist and occasionally collide. Parents' information world emphasizes the cultural and linguistic nuances of Korean, moral intentionality, and structured learning. In contrast, teens' information world prioritizes English fluency, speed, and adaptability—values aligned with the structure of Western educational institutions and AI system affordances.

In this context, AI becomes a boundary object (Star & Griesemer, 1989): accessible across generations yet interpreted through divergent lenses. Its design—optimized for English and often insensitive to cultural

context—structurally privileges teens' epistemic frameworks while marginalizing parents'. As noted in P1's critique of "awkward" Korean translations, these design choices are not just usability flaws but material manifestations of deeper power imbalances. Such mismatches reinforce parental perceptions of exclusion, making AI literacy a site of intergenerational struggle rather than seamless adoption.

## 5.3. Linguistic Inequity and Epistemic Injustice

This asymmetry reflects broader structural biases in AI development. Our data underscore the risks of linguistic marginalization, as teens gravitate toward English for its perceived clarity and utility, while parents encounter friction and alienation in navigating Korean interfaces. This supports prior work on epistemic injustice (Fricker, 2007), where the knowledge systems and linguistic preferences of minority users are undervalued or mistranslated.

T7 and T9's comments about defaulting to English, contrasted with P7's feeling of being "locked out," reveal how language becomes a fault line within AI-mediated family life. These dynamics exacerbate intergenerational divides, further marginalizing the cultural capital that parents bring to digital interactions. Over time, this may contribute to cultural erosion and linguistic displacement, particularly among second-generation users.

## 5.4. Co-Construction of AI Literacy: Dialogic Learning and Shared Authority

Despite these tensions, our findings also highlight opportunities for intergenerational co-learning. As seen in teens' role as digital intermediaries and parents' ethical scaffolding, families engaged in informal but substantive knowledge exchanges. These interactions were often experiential rather than instructional—e.g., troubleshooting AI outputs together, reflecting on translation errors, or debating answer credibility.

This co-construction reflects principles of Distributed Cognition (Pea, 1993) and interdependent navigation (Seo & Ammari, 2025), where literacy is dispersed across people, tools, and cultural scripts. Rather than a top-down flow of expertise, AI literacy emerged as a shared practice that evolves through everyday collaboration, inquiry, and negotiation. T2's remark about learning from her mother's questions underscores this dialogic structure, challenging assumptions that technological expertise is a one-way transfer from youth to elders.

## 5.5. Toward Justice-Oriented AI Design and Literacy

Our study calls for rethinking both design and education. First, culturally inclusive AI must move beyond translation toward culturalization—incorporating diverse linguistic norms, values, and interaction styles into system design. Second, AI interfaces should support interdependent use, offering affordances for shared reflection (e.g., annotation layers, family discussion prompts) rather than targeting individual users alone.

Third, AI literacy curricula must evolve beyond functional skills to include critical reflection on algorithmic bias, data politics, and cultural representation. For immigrant families, this means creating spaces where their ways of knowing are validated, not overwritten. By reframing AI literacy as a relational, co-constructed practice, we can design tools and systems that reflect the lived complexities of multicultural households.

Ultimately, the Korean immigrant families in our study show that AI literacy is not only about mastering tools—it is about navigating values, language, and power in the digital age. Their experiences demand that we recenter justice, not just usability, in how we design and teach AI for diverse communities.

# 6. Limitations and Future Directions

While the specificity of our sample—educationally privileged Korean immigrants in the U.S.—is a limitation to generalizability, it is also a strategic strength. This "critical case" allowed us to isolate and magnify the dynamics of cultural and linguistic friction that might be obscured in a broader sample. It demonstrates that even for a relatively resourced immigrant group, significant barriers to equitable AI engagement persist.

Future research must expand this inquiry. How do these dynamics play out in immigrant communities with different cultural values regarding education and authority? How do class and documentation status intersect with these intergenerational negotiations? Most importantly, we need participatory design research, co-creating AI systems with multilingual families, not just for them, to build a truly inclusive and equitable technological future.

# 7. Conclusion

This study examined how Korean immigrant families navigate AI technologies in their everyday lives, revealing that AI literacy is neither a fixed skillset nor an individual accomplishment. Instead, it is a relational and negotiated practice—shaped by cultural values, linguistic fluency, intergenerational trust, and the structural affordances and limitations of AI systems themselves. Through the dual lenses of *interpretive gatekeeping* and *convenient critical deferment*, we showed how parents and teens collaboratively but asymmetrically engage with AI, balancing ethical caution with pragmatic utility.

Our findings reconceptualize AI literacy as co-constructed across generations, embedded within family dynamics, and inseparable from questions of language, identity, and power. They highlight that trust in AI is deeply conditioned by cultural epistemologies; that fluency in English opens up technological possibilities while simultaneously marginalizing non-English speakers; and that role reversals, where teens tutor parents, often coexist with more traditional forms of parental moral guidance.

This research urges scholars and designers to resist universalist assumptions in AI literacy frameworks and to foreground the lived, culturally situated realities of immigrant families. As algorithmic systems become more central to daily life, so too does the need for justice-centered design approaches that value pluralism, dialogic learning, and culturally responsive interaction. We call for AI systems that do not merely accommodate linguistic diversity, but respect it; that do not simply empower individuals, but support interdependent, relational use; and that recognize immigrant families not as technological outsiders, but as critical co-creators of AI futures.